\documentclass{aastex62}

\usepackage{amsmath}
\usepackage{epsfig}
\usepackage{amssymb}
\usepackage{graphicx}
\usepackage{color}
\usepackage{natbib}

\newcommand{\Msol}{$M_{\odot}$}

\received{November 1, 2020}

\accepted{January 22, 2021, by ApJ}

\shorttitle{DCBHs with JWST}
\shortauthors{Reg{\H o}s, Vink\'o \& Stermeczky}

\begin{document}

\title{Detection of Tidal Disruption Events around Direct Collapse Black Holes at High Redshifts \\ with the {\it James Webb Space Telescope}}

\correspondingauthor{Enik\H o Reg\H os}
\email{Enikoe.Regoes@gmail.com}

\author[0000-0002-9498-4957]{Enik\H o Reg\H os}
\affil{ Konkoly Observatory,  CSFK, 
Konkoly-Thege M. \'ut 15-17,
Budapest, 1121, Hungary}

\author[0000-0001-8764-7832]{J\'ozsef Vink\'o}
\affil{ Konkoly Observatory,  CSFK, Konkoly-Thege M. \'ut 15-17, 
Budapest, 1121, Hungary}
\affil{ELTE E\"otv\"os Lor\'and University, Institute of Physics, P\'azm\'any P\'eter s\'et\'any 1/A, Budapest, 1117 Hungary}
\affil{Department of Optics \& Quantum Electronics, University of Szeged, D\'om t\'er 9, Szeged, 6720, Hungary}

\author{Zs\'ofia V. Stermeczky}
\affil{ELTE E\"otv\"os Lor\'and University, Department of Astronomy, P\'azm\'any P\'eter s\'et\'any 1/A, Budapest, 1117 Hungary}

\begin{abstract}
This is the third sequel in a series discussing the discovery of various types of extragalactic transients with the {\it James Webb Space Telescope} in a narrow-field ($\sim 0.1$ deg$^2$), moderately deep ($m_{AB} \sim 27$ mag) survey. In this part we focus on the detectability and observational characteristics of Direct Collapse Black Holes (DCBH) and Tidal Disruption Events (TDE) around them. We use existing models for DCBH accretion luminosities and spectra as well as for TDE light curves, and find that accreting DCBH seeds may be bright enough for detection up to $z \sim 7$ with JWST NIRCam imaging, TDEs of massive ($M \gtrsim 50$ \Msol) stars around them can enhance the chance for discovering them as transient objects, { although the rates of such events is low, a few per survey time.} TDEs around non-accreting black holes of $M \sim 10^6$ \Msol\ may also be detected at $z < 7$ redshifts in the redder NIRCam bands between 3 and 5 microns. It is also shown that accreting DCBHs appear separate from supernovae (SNe) on the NIRCam color-color plot, but TDEs { from quiescent black holes} fall in nearly the same color range as Superluminous Supernovae (SLSNe), which makes them more difficult to identify.
\end{abstract}


\section{Introduction} \label{sec:intro}

By the end of the cosmic dark ages naissent stars and galaxies initiate the reionization of the intergalactic medium. 

The First Light At Reionization Epoch (FLARE) survey \citep{flare} project was set up to transform the search for new objects appearing during the reionization epoch to the time domain. It aims to find supernovae and direct collapse black holes with the {\it James Webb Space Telescope} (JWST). 
FLARE proposes to find and characterize the first stars in the Universe in a shallow survey down to about 27 AB-magnitude.
It uses a discovery space of transient phenomena with JWST in the early Universe.
The proposed survey will find transients to reveal the state of the Universe when the first stars and black holes were formed. 
See \citet{flare} for more details. 

\citet{rv19} (RV) examined the supernovae to be discovered with JWST in a 0.1 square degree field survey
in 3 years, corresponding to the original survey science goals set out in \citet{flare}. 
RV examined the detection, brightness and expected number counts of superluminous supernovae (SLSNe) and Type Ia supernovae
(SNe Ia) in the FLARE survey.

In a follow-up paper \citet{rvz20} examined the possibility for identifying pair-instability supernovae (PISN) in the FLARE survey. 
As Population III stars are short-lived, their rapid enrichment hides Pop III star formation, and even deep JWST exposures will tend to see metal-enriched Population II stars.
To constrain the high-mass end of stellar mass functions, the search for PISNe would reveal if star formation led to masses in excess of 150 \Msol, the threshold for the onset of the pair-creation instability. \citet{rvz20} found that individual PISN events are bright enough to be detected with JWST, but the challenge is their low surface density. 

In the present paper, which is the third in a series dealing with discovery of transients with JWST, we study accretion onto black holes formed by direct collapse from the primordial gas. 
It is expected that JWST can detect these most luminous and earliest cosmic messengers in a sufficiently wide field. 
Such a survey is vital in addressing the key scientific goals of JWST. 

Recently \citet{flare} reviewed the theoretical background of Direct Collapse Black Hole (DCBH) formation and their detectability with JWST. Here we briefly repeat the most important points for completeness.

DCBHs that are thought to be the origin of the first supermassive black holes (SMBH) could be formed via two channels: either from the direct collapse of the high-redshift primordial gas halos \citep[e.g.][]{bl03}, or from the core collapse of the first stars \citep{whalen12}. 
The primordial initial mass function was rather broad, extending to possibly very high masses of Pop~III stars. This would imply that massive black hole remnants are quite common.

The formation of the first SMBH seeds took place at $z \gtrsim 10$ , i.e. when the age of the Universe was $t \lesssim 500$ Myr. These SMBH seeds made important contribution to the growth of early ($z \sim 7$) SMBHs \citep{pacucci17}.

The observational characteristics and JWST detectability of DCBHs were studied by \citet{nata17}, who showed that accreting DCBHs can be as bright as $m_{AB} \sim 26$ mag in the infrared, thus, detectable with JWST.
{ \citet{visbal18} propose that future X-ray missions like Lynx, combined with infrared observations, could distinguish high-redshift DCBHs through their small
host galaxies, that is the high BH mass to stellar mass ratios of the faintest observed quasars.}


Comparison between the stellar spectral energy distribution (SED) of some GOODS-S objects \citep{illing16} with computed SEDs of black holes grown of DCBHs show good agreement. These objects are characterized by their higher infrared color indices, i.e. their infrared SEDs increase steeply toward longer wavelengths, as predicted for DCBHs.
The steepness of the SED is also their possible observational signature in addition to infrared magnitudes. 

Observations of transients provide another way to characterize the early Universe. Tidal disruption events (TDEs) and the intrinsic variability of accretion flow can trace the build-up of billion solar mass SMBHs within a few hundred millions of years. 
{ \citet{alex17} derive a minimal mass of $3\times 10^5$ \Msol 
for present-day central black holes in galaxies, and point out that the lack of intermediate-mass black holes at low redshifts has observable implications for tidal disruptions.}

Adopting event rates from the literature, \citet{fialkov17}
established trends in the redshift evolution of the TDE number counts and their observable signals.
{ They find TDE rates that are weak functions of redshift. 
On the other hand, the redshift evolution of the TDE rate is very uncertain, because the dominant mechanism
for loss-cone feeding is unknown \citep[see][for a recent review]{stone20}.
}

While TDEs have a low rate relative to other transient events, their discovery rate was augmented by the { 1 -- 5 day} cadence optical surveys as the Palomar Transient Factory \citep{law09} , 
the All-Sky Automated Search for Supernovae (ASAS-SN) { \citep{holo19}},
{ the Sloan Digital Sky Survey \citep{velzen11}
and the Zwicky Transient Facility \citep{velzen20}.}
Pan-STARRS found a { number of important} TDEs, { e.g. the famous optical PS1-10jh \citep{monte13, gui14, bogda14, gezari15, strubbe15} }.
ZTF contributed to a sample useful for statistical analysis, and the forthcoming { Legacy Survey of Space and Time (LSST) by the Vera C. Rubin Observatory}
will also detect many events per year to enable it.

Supernovae in galactic nuclei may confuse optical surveys for tidal disruption events: \citet{sq11}
estimate that nuclear Type Ia supernovae are two orders of magnitude more common than TDEs at z $\sim$ 0.1 for ground-based surveys.
Nuclear Type II SNe occur at a comparable rate but can be excluded by pre-selecting red galaxies.
The contamination from SNe can be reduced by high-resolution follow-up imaging with adaptive optics or HST.
Predictions help transient surveys on their potential for discovering TDEs. Detecting and characterizing first SLSNe and TDEs are goals of the FLARE project for JWST \citep{flare}. 

Section 2 discusses Direct Collapse Black Holes and their Tidal Disruption Events,
TDE theory, the distribution of black hole masses and the expected TDE rates in the FLARE survey.
Section 3 describes simulations of TDE light curves, the models superimposed 
on DCBHs and their detection in various JWST NIRCAM filters.
Section 4 presents the results of the simulations and their predicted ranges on JWST color -- color plots. We conclude our results in Section 5.

\section{Direct Collapse Black Holes and their Tidal Disruption Events} \label{dcbh}

{ \subsection{DCBH Theory Overview} }

N-body simulations of the large-scale structure are used to estimate the spatial density of observable SMBHs.
The large range of predictions comes from uncertainties in the critical Lyman - Werner radiation field that can suppress H$_2$ formation, from clustering of DCBH formation sites and feedback.
\citet{habou16} summarize these models. They derive a number density that is also constrained by the observed cosmic near-infrared background fluctuation level.

Compton-thick DCBHs are more detectable as more energy is reprocessed to rest-frame UV/optical, and redshifted to near-infrared bands \citep[e.g.][]{yue14}. 

There are two seeding model families for DCBHs: light seed black holes are the remnants of Pop III stars, while heavy seeds are from the direct collapse of gas clouds.
Several models exist for the accretion history as well: sub-Eddington accretion, 
slim disk models and torque limited growth models \citep{benami18}. 

There is evidence for a direct collapse black hole in the 
Lyman~$\alpha$ source CR7 \citep{smith16}. 
The CR7 galaxy at $z = 6.6$ has a combination of exceptionally bright Ly~$\alpha$ and He~II 1640 \AA\ line emission but absence of metal lines.
As a result, CR7 may be a candidate host of a DCBH.
A massive black hole with a non-thermal Compton-thick spectrum reproduces its Ly~$\alpha$ signatures.

The DCBH scenario describes the isothermal collapse of a pristine gas cloud directly into a massive, $10^4$ -- $10^6$ \Msol\ black hole.
Large HI column densities of primordial gas at $10^4$ K temperature provide conditions for the pumping of the 2p-level of atomic hydrogen by trapped Ly~$\alpha$ photons.
This gives rise to stimulated fine-structure maser emission at 3.04 cm. 
Detection of the redshifted 3-cm emission line could provide direct evidence for the DCBH scenario \citep{dijk16}.

Formation of binary and multiple DCBHs in atomically cooled halos increases the possibility of detecting tidal disruption events (TDE) in the near infrared with JWST. 
The effect of binary stellar populations on DCBH formation through the irradiating Lyman -- Werner field was also considered (\citet{latif}).

\citet{kashi16} give detailed analytical calculations for the formation of a nuclear cluster around DCBHs and the TDE of some of the cluster members scattered within the tidal radius of the central black hole. They find that such events might be detectable in X-rays up to $z \lesssim 20$, and the afterglow may also be visible with radio telescopes or JWST. 

\subsection{Distribution of black hole masses}

Formation of a broad distribution of clustered primordial black holes is predicted e.g.
from Higgs inflation (dilaton) models, which could constitute today's dark matter.
In less exotic models, light seed black holes result from remnants of Pop III stars, while heavy seeds are formed from the direct collapse of gas clouds
(\citet{ferrara14}).

The lognormal distribution for the birth mass function of Intermediate Mass ($10^4$ -- $10^6$ \Msol) DCBH seeds gives a tapered power law for the mass function
with limiting distribution of modified lognormal power law 
\citep{basu19} as
\begin{equation}
\frac{dn}{d \ln M} = \frac{\alpha}{2} \exp\left(\alpha\mu_0 + \alpha^2\sigma_0^2/2\right)M^{-\alpha} \times {\rm erfc} \left(  \frac{1}{\sqrt{2}}(\alpha\sigma_0-\frac{\ln M-\mu_0}{\sigma_0 })\right),
\label{eq-taperedpl}
\end{equation}
with infinite time for the creation of all DCBHs,
where $\mu_0$ and $\sigma_0$ are the mean and the standard deviation of the logarithmic mass distribution of the DCBH seeds. \citet{ferrara14} gives 
$\mu_0=11.7$ (corresponding to peak at mass of $10^{5.1}$ \Msol) and  $\sigma_0=1.0$.  

The break in the power law is 
a marker of the end of the DCBH growth era.
$\alpha=0.5$ with break-point related parameter $\beta=8.4$ (in the tapered power law) fit the quasar 
luminosity function as well and reveals Direct Collapse Black Hole growth theory.



The formation of DCBH seeds took place between $20 < z < 12$ \citep{yue14}. 
The increase of the DCBH number density, characterized by the DCBH growth rate $\lambda = d \log n_{\rm DCBH} / d \log t$, was found as $\lambda \sim 27.7$ Gyr$^{-1}$ if the critical flux of the Lyman-Werner radiation was $J_{crit} = 300 \times 10^{-21}$ erg~s$^{-1}$~cm$^{-2}$~Hz$^{-1}$~sr$^{-1}$ \citep{dijk14, basu19}.  The growth of each DCBH is thought to happen exponentially, close to the Eddington-rate, but periods of super-Eddington growth are also possible \citep{pacucci17, basu19}.   




\subsection{Tidal Disruption Event theory and modeling}

The bases of TDE dynamics have been laid down by \citet{rees88} and \citet{ek89}, among others. Stars in galactic nuclei can be captured or tidally disrupted by a central black hole. 
Some debris would be ejected at high speed, the remainder would be swallowed by the hole, causing a bright flare lasting at most a few years. This results in a light curve characterized by the famous $t^{-5/3}$ decline rate, as predicted by { \citet{rees88} and} \citet{phin89}.


As the TDE accretion rates are super-Eddington, which can last a year, the peak luminosity can be as bright as the brightest superluminous supernovae. 
Their peak luminosities combined with their characteristic light curve shapes can be used to classify transient events as TDEs.

\citet{sq09} highlight some of the observational challenges associated with studying tidal disruption 
events in the optical. 
\citet{lr11} { predict} that after a few months TDE optical and UV light curves scale as $t^{-5/12}$, and are thus substantially 
flatter than the well known $t^{-5/3}$ decline.
The X-ray band, instead, is the best place to detect the $t^{-5/3}$ behaviour, although only for roughly a year, before the emission steepens exponentially.
{ The observed properties of optical TDEs \citep{komossa15, velzen20} show that the 5/12 power-law is
not observed during the first 1 -- 3 year of the light curve. Instead, \citet{velzen20} found that the observed optical light curves of a sample of 33 TDEs can be described by a power-law with an average index of $p \sim -1.65 \pm 0.65$, which is remarkably close to the canonical $p= -5/3$ value, albeit with significant event-to-event scatter. }

{ There are other observational indications which suggest that, in reality, TDEs can be more diverse than it was suspected based on the first, simplified  theoretical picture. For example, 
ASASSN-14li was a particularly well-observed TDE having extensive multi-wavelength data, which can be modeled self-consistently by the disruption of a $\sim 1$ $M_\odot$ star around a $\sim 10^{6.5} ~M_\odot$ SMBH \citep{krolik16}.
\citet{holo16} found that the early pseudo-bolometric light curve { is most consistently fit} by an exponential decay of $e^{-t/60}$ { (in days)}, while the $t^{-5/3}$ power-law gives { moderately} worse fit.
{ The early-time light curve can be explained by both a power law and an exponential decay.}
More recent simulations \citep{law-smith19} as well as semi-analytical models \citep{krolik20} confirmed that real TDEs can be much more complex than the simple analytical models suggest. While simple models might be capable of giving order-of-magnitude estimates for the basic parameters (e.g. masses), more detailed analyses would probably require simulations. This caveat should be kept in mind while considering simple TDE model light curves.  }









\subsection{Expected { DCBH and} TDE Rates in FLARE}

The surface number density of DCBHs brighter than 26.5 mag at 4.5 $\mu$ in \citet{flare}, predicted from various models \citep{aga12, yue13, dijk14},
shows large variations, { ranging from $10^{-10}$ to $10^{-1}$ Mpc$^{-3}$}.
The large range comes from uncertainties in the critical Lyman - Werner radiation field to suppress H$_2$ formation, clustering of DCBH formation sites and feedback.
{ However, \citet{habou16} display that the more modern simulations corresponding 
to large cosmological volumes and sophisticated modelling of physical processes
give the highest number densities of DCBHs, and the variation in the above listed
3 models is only 2 orders of magnitude.}
Since the comoving number density of DCBHs { in the \citet{yue13} models is as high as} $n_{DCBH} \sim 0.1$ Mpc$^{-3}$ at $z=8-10$, 
{ we adopt this value (also as upper limit) for our rate estimates.}. 
Normalizing to the observed near-infrared background
fluctuations it can be even 0.1 Mpc$^{-3}$ at $z=13$ \citep{yue13}.
 

In \citet{kashi16}
the TDE rate is $\sim 10$ per DCBH within 1 Myr at the DCBH early growth stage.
{ Their 
rate applies to TDEs from $M \sim ~\mathrm{few} \times 10 - 100$ \Msol\ stars, their effective characteristic mass being higher than 40 \Msol.}

If the TDE happens through the whole lifetime of a DCBH, then
multiplying the { $n_{DCBH} \sim 0.1$ Mpc$^{-3}$} surface number density of a steadily accreting DCBH { by the $10 \cdot T_{DCBH} /(1~\mathrm{Myr})$ events/DCBH rate given above, where $T_{DCBH}$ is the expected lifetime of the accreting DCBH,} we obtain the detectability of TDEs in the FLARE survey as
$n_{TDE} \sim 0.5$ deg$^{-2}$yr$^{-1}$ per unit redshift bin { (see also in \citet{flare} )}. 
Thus, we can expect $\sim 1 - 2$ events in the entire redshift range during the total survey time ($\sim 3$ years).
If this is enhanced by normalizing to NIR background fluctuations, the expected TDE counts in the proposed FLARE survey are quite reasonable.
{ Note that if the final field of view of the FLARE survey is increased up to 0.8 square degree, rather than 0.1, the TDE counts in the wide field survey will be higher by a factor of 8, i.e. about $\sim 10$ events.}


\section{Simulations}\label{sec:simulations}

In this paper we use existing models for DCBH accretion and TDE light curves to estimate the observability of such events with JWST during the FLARE survey. Note that the final parameters of the FLARE survey are not yet fixed, but even an area of 0.1 square degree provides valuable numbers. 

While the applied filters are also to be finalized, for present simulation we adopt the same 4 NIRCam filters (F150W, F200W, F356W and F444W) as used in \citet{rv19} and \citet{rvz20}. 

As in the previous papers, we apply the {\tt sncosmo} code \citep{sncosmo} for calculating the observed signals of the redshifted DCBH/TDE events. 

\subsection{DCBH models}\label{sec:dcbhmodel}

We adopt the models of \citet{pacucci15} for the SEDs of DCBHs starting at $10^5$ \Msol\ mass and growing as a function of time up to $10^6$ - $10^7$ \Msol. Since the timescale of DCBH growth is $\sim 10^7$ years, we consider only those epochs when the model fluxes are the highest, providing the best-case scenario for the observability of these events. For the accretion process we consider both the standard and slim-disk accretion models as given by \citet{pacucci15}. 

\subsection{Modeling TDE light curves}\label{sec:tdemodel}

\begin{table}
    \centering
    \begin{tabular}{c|c}
        \hline
        Parameter & Values  \\
        \hline
        $M_{BH}$ & $10^6$ M$_\odot$ \\
        $M_*$ & 10, 50, 200 M$_\odot$ \\
        $R_*$ & 5.6, 18.8, 53.2 R$_\odot$ \\
        $\beta$ & 1.0 \\
        $\eta$ & 0.1 \\
        $f_{out}$ & 0.3 \\
        $f_v$ & 2.0 \\
        $i$ & $0^o$ \\
        $z$ & 5,6,7,8,9 \\
        \hline
    \end{tabular}
    \caption{Parameters for simulated TDE light curves (see text).}
    \label{tab:tde-params}
\end{table}

For computing the temporal evolution of the Spectral Energy Distribution (SED) of a TDE, we use the parameterized model described by \citet{lr11}. We assume that stars approach the supermassive BH on parabolic orbits, and the closest encounter occurs at the tidal radius $R_T = R_* (M_{BH} / M_{*})^{1/3}$, i.e. the impact parameter $\beta = R_T / R_p = 1$, where $R_p$ is the pericenter distance \citep{lr11}. 
{ We do not consider partial disruptions having $\beta < 1$ \citep[e.g.][]{grr13}.}

After disruption, half of the stellar material is thought to leave the system, while the other half that were closer to the BH at the moment of the pericenter passage is assumed to fall back to onto the BH via super-Eddington accretion. The accretion luminosity produces the initial flare that is usually referred to as the super-Eddington wind. A fraction of the fallback material leaves the system due to the wind, while the remaining fraction forms an accretion disk that keeps the material accreting to the BH. 

Following \citet{lr11}, the following parameters are applied to describe the TDE model SED \citep[see][for more detailed description]{lr11}:
\begin{itemize}
    \item{mass of the BH ($M_{BH}$, in $10^6$ M$_\odot$ units)},
    \item{mass of the disrupted star ($M_*$, in M$_\odot$ units)},
    \item{radius of the disrupted star ($R_*$, in R$_\odot$ units)},
    \item{impact parameter ($\beta = R_T/R_p = 1$)},
    \item{radiation efficiency of the accretion ($\eta = 0.1$, in units of the mass accretion rate)},
    \item{wind mass fraction ($f_{out} = 0.3$, in units of the fallback mass)},
    \item{wind velocity factor ($f_v$ = 2, in units of the escape velocity from the disk)},
    \item{disk inclination angle ($i = 0$ deg, assuming a face-on disk)}.
\end{itemize}

{ We assume that the disrupted stars are zero-metallicity Population~III objects during their main sequence (MS) phase. The models by \citet{ohkubo09} predict $R_* \sim M_*^{0.75}$ mass-radius relationship for such objects, which is adopted in our calculations. 

Note that the \citet{lr11} model predicts a shallower ($\sim -5/12$) power-law decline for the optical light curve than the canonical $-5/3$ value supported by the observations (see Section~2.2). This caveat limits the applicability of this model for analyzing the observed events. On the other hand, in this paper we are primarily concerned with {\it detections} of TDEs, where the peak luminosities (related to mass fallback rates and radiation conversion efficiency) are more important than the actual shapes of the light curves. In this respect the applied model gives reasonable predictions.   

The \citet{lr11} model assumes a simple constant mass-energy distribution for the disrupted star. In fact, based on semi-analytical calculations and simulations, \citet{lkp09} showed that more realistic density distributions affect the TDE light curves mainly during the pre-maximum phases, causing a more gradual rise of the light curve to maximum. The peak mass accretion rate, hence the peak brightness, is only mildly affected and stays within a factor of $\sim 2$ for models with polytropic index in between $\gamma = 4/3$ and 1.8 \citep[see Fig.10 in][]{lkp09}. Similar results are presented by \citet{grr13}. Therefore, the light curves presented in Section~\ref{sec:results} would not change drastically when other forms of density distribution were assumed.}

{ The structure of Pop III stars is not very different from the structure of Population I stars, although they are more compact as the opacity is lower.
Therefore it is harder to disrupt them and their debris is closer to the black hole. This smaller orbit will result in a shorter time scale in their light curve.
Even if the LC model used is not accurate (or match known optical TDEs at low
redshift) it may be suitable to examine their behaviour at high redshift and their detection by JWST.}

{TDEs in an accreting black hole will likely have different observational
signature compared to TDEs from quiescent black holes \citep{chan20}, but this is beyond the scope of this work.}

\subsection{TDEs around DCBHs}

Monochromatic TDE light curves (LCs) at rest-frame frequencies corresponding to the central wavelengths of JWST/NIRCam F150W, F200W, F356W and F440W filters in the observer's frame are calculated via the formulae given by \citet{lr11}. Contributions from both the wind and the disk are considered. Then, the calculated TDE fluxes are added to the model SEDs of accreting DCBHs adopting the models of \citet{pacucci15}. Both the standard and slim-disk accretion models are considered, resulting in slightly different final combined DCBH-TDE model LCs.  Finally, the rest-frame model SEDs are scaled to different redshifts ($z$) and the corresponding luminosity distances ($D_L$, in Mpc) assuming $\Lambda$-CDM cosmology with the following parameters:  \(\Omega_m=0.315, \Omega_\Lambda=0.685, H_0=67.4 \) km/s/Mpc \citep{planck18}, applying the {\tt astropy.cosmology} module in Python \citep{2013A&A...558A..33A}.

The results of these calculations are presented and discussed in the following section.

\section{Results and discussion}\label{sec:results}

\begin{figure*}
\centering
\includegraphics[width=18cm]{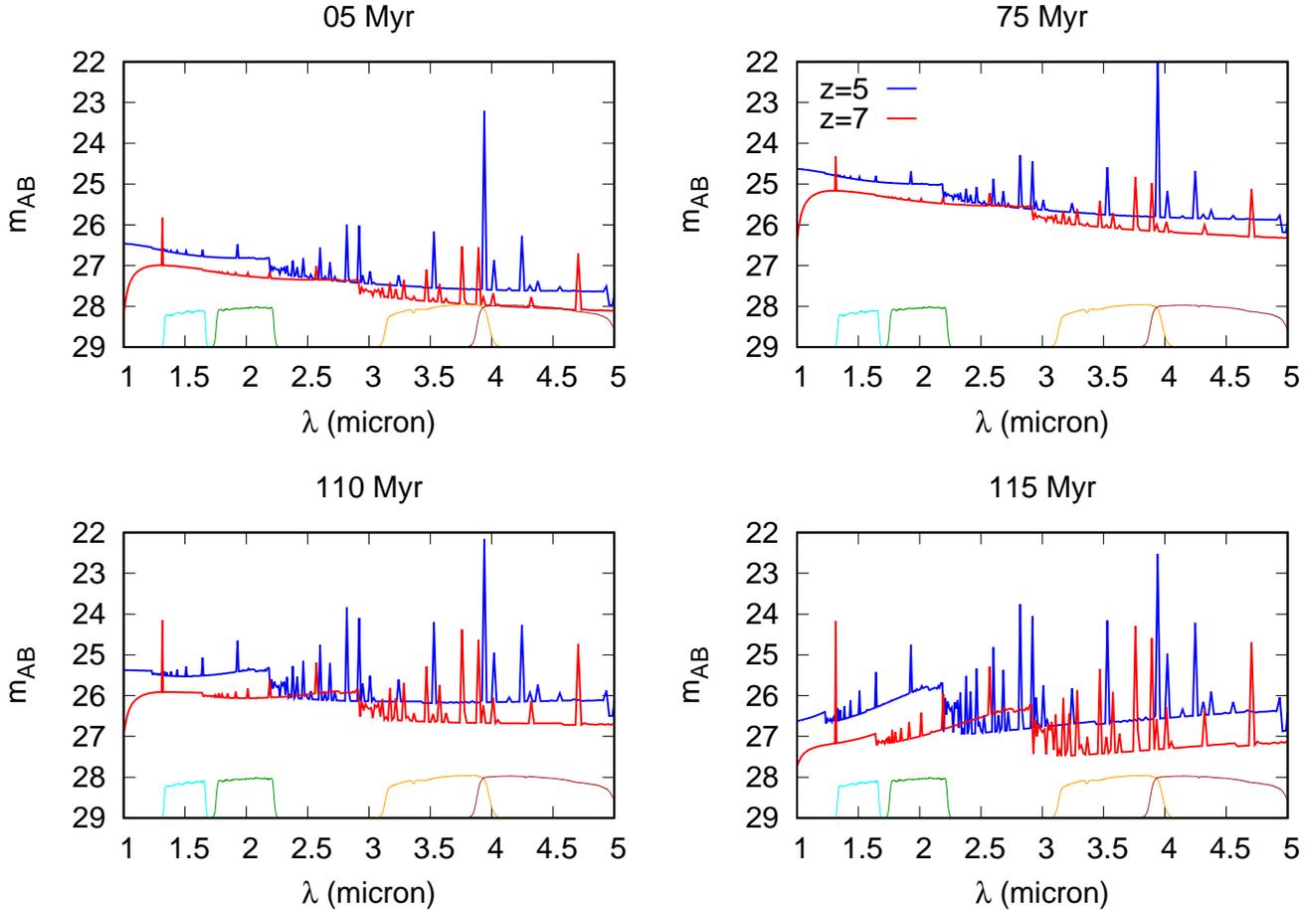}
\caption{Models of DCBH spectra \citep{pacucci15}, redshifted to $z=5$ (blue curve) and 7 (red curve), assuming standard accretion at four different ages after BH formation. The age of the spectra are indicated in the title of each subpanel. The colored curves in the bottom of each panel represent the bandpass functions of the considered JWST/NIRCam filters (F150W, F200W, F356W and F444W, respectively). Fluxes above $m_{AB} \sim 27$ mag are expected to be detectable on NIRCam frames during the FLARE survey \citep{flare}}.
\label{fig:dcbh_std_sp}
\end{figure*}

\begin{figure*}
\centering
\includegraphics[width=18cm]{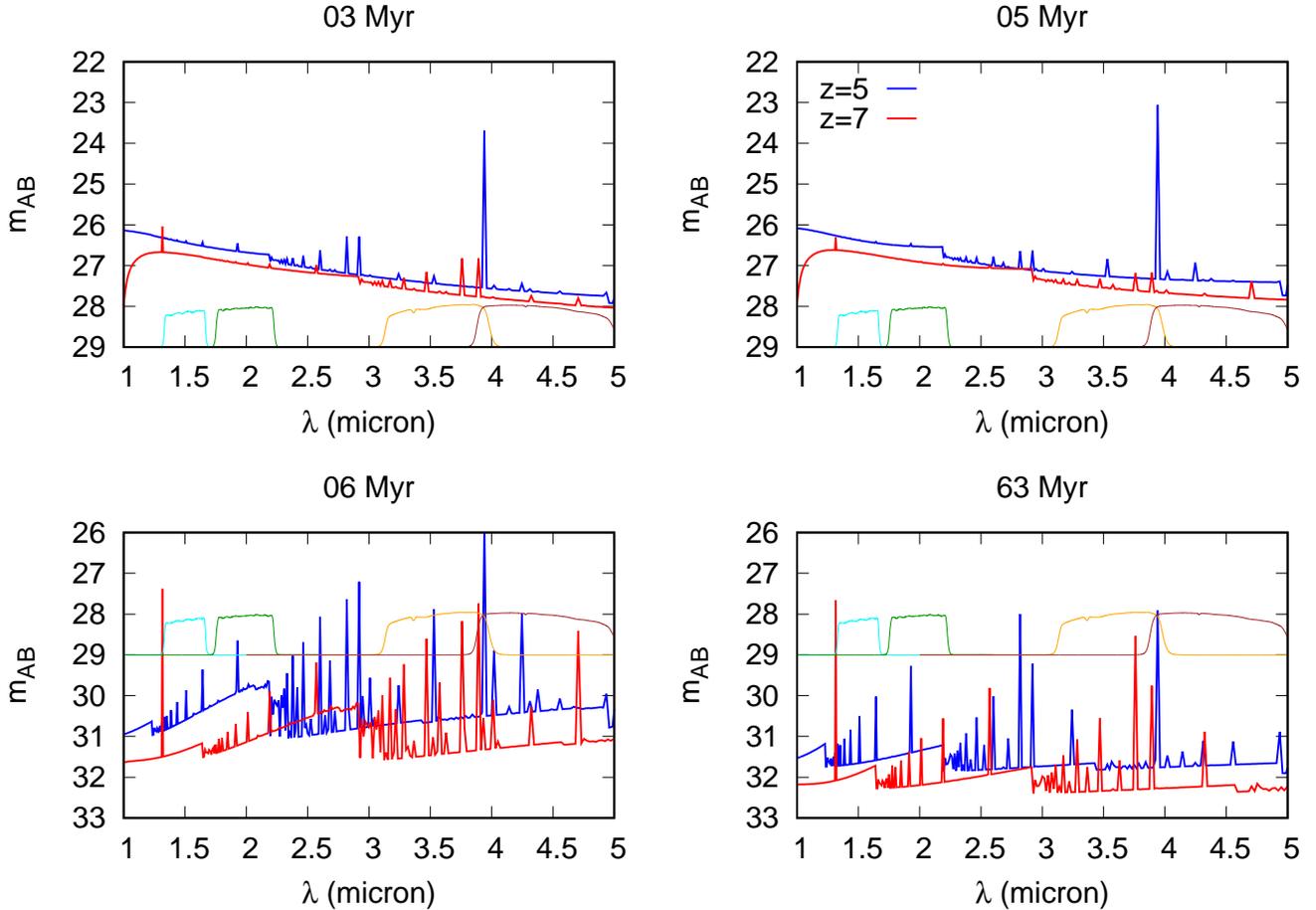}
\caption{The same as Figure~\ref{fig:dcbh_std_sp} but for accretion via slim disk. }
\label{fig:dcbh_slimd_sp}
\end{figure*}

Figures~\ref{fig:dcbh_std_sp} and \ref{fig:dcbh_slimd_sp} show the DCBH models of \citet{pacucci15} assuming standard and slim-disk accretion, respectively. The models are redshifted to $z=5$ and 7 to illustrate the observer-frame spectra of DCBHs with different ages after formation. The bandpasses of the four JWST/NIRCam filters are also plotted. It is seen that DCBHs with standard accretion can be detectable above $m_{AB} \sim 27$ AB-magnitude in each NIRCam bandpass up to $z \sim 7$ \citep[see also][]{pacucci15}.   

%

\begin{figure*}
\centering
\includegraphics{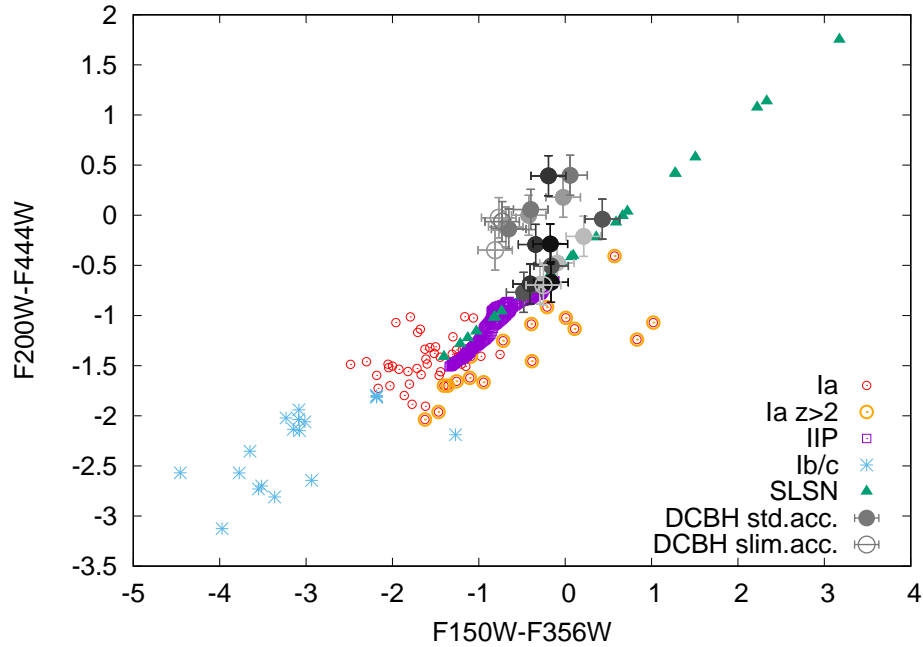}
\caption{Color-color plot for NIRcam filters showing various types of transients. Grey circles are DCBHs from both standard and slim-disk accretion models, while the colored symbols illustrate the expected positions of several types of supernovae \citep[see][]{rv19}.}
\label{fig:dcbh_cc}
\end{figure*}

In Figure~\ref{fig:dcbh_cc} the expected positions of DCBHs at various ages and redshifts (between $4 < z < 9$) are plotted on the JWST/NIRCam color-color plot. The colored symbols illustrate the positions of various types of supernovae as shown in \citet{rv19}. As seen, DCBHs are tightly clustered around 0 in both colors, redward above the diagonal sequence populated by supernovae, regardless of age, redshift or accretion model. 

\begin{figure*}
    \centering
    \includegraphics[width=18cm]{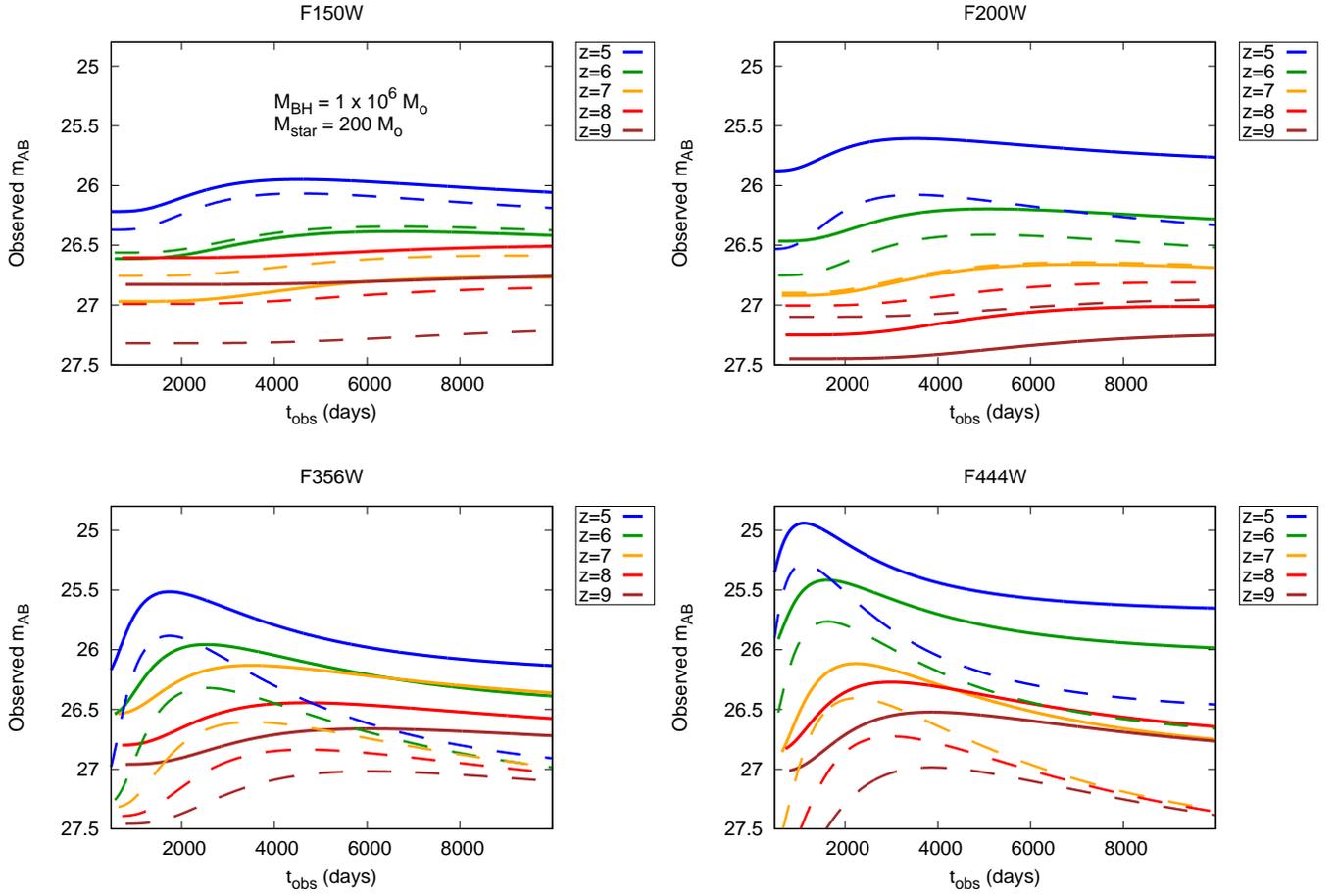}
    \caption{Light curves of a Tidal Disruption Event in the JWST/NIRCam filter wavelengths at various redshifts, superimposed on the accretion SEDs of DCBHs. The mass of the DCBH is $10^6$ \Msol and the stellar mass is 200 \Msol. Solid curves denote models with standard accretion, while dashed curves correspond to models with slim-disk accretion \citep{pacucci15}.}
    \label{fig:tde6200}
\end{figure*}

Figure~\ref{fig:tde6200} shows light curves of a TDE from a 200 \Msol\ star around a DCBH of $10^6$ \Msol\ in the considered JWST/NIRCam filter wavelengths at various redshifts (color-coded in the legend). Accretion luminosity onto the DCBH is added to the predicted TDE fluxes as described in \S\ref{sec:tdemodel}. 
Solid lines indicate standard DCBH accretion at 110 Myr, while dashed lines show the models with slim-disk accretion at 0.5 Myr \citep{pacucci15}.

\begin{figure*}
    \centering
    \includegraphics[width=18cm]{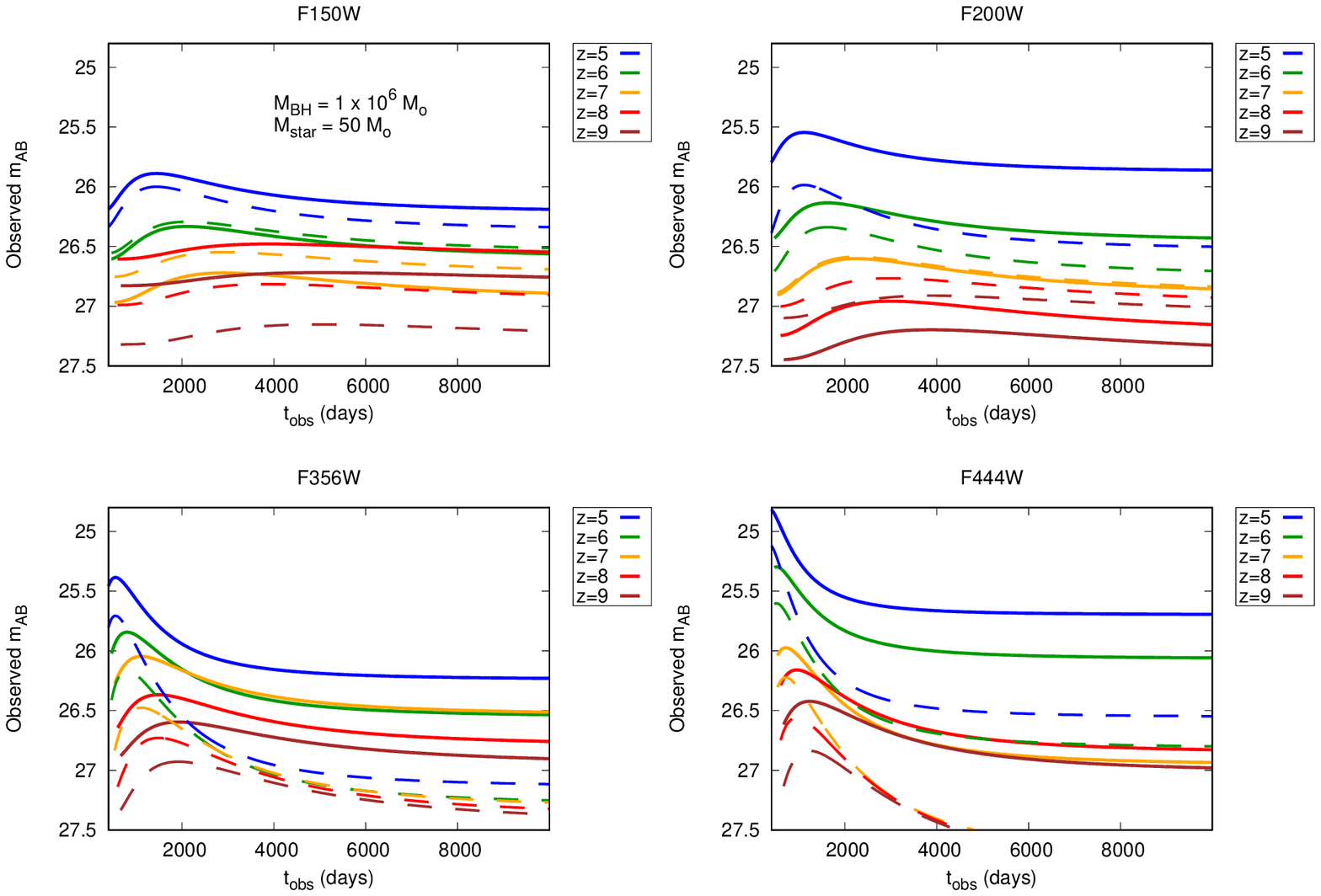}
    \caption{The same as Fig.\ref{fig:tde6200} but for $M_{star} = 50$ M$_\odot$.}
    \label{fig:tde650}
\end{figure*}

\begin{figure*}
    \centering
    \includegraphics[width=18cm]{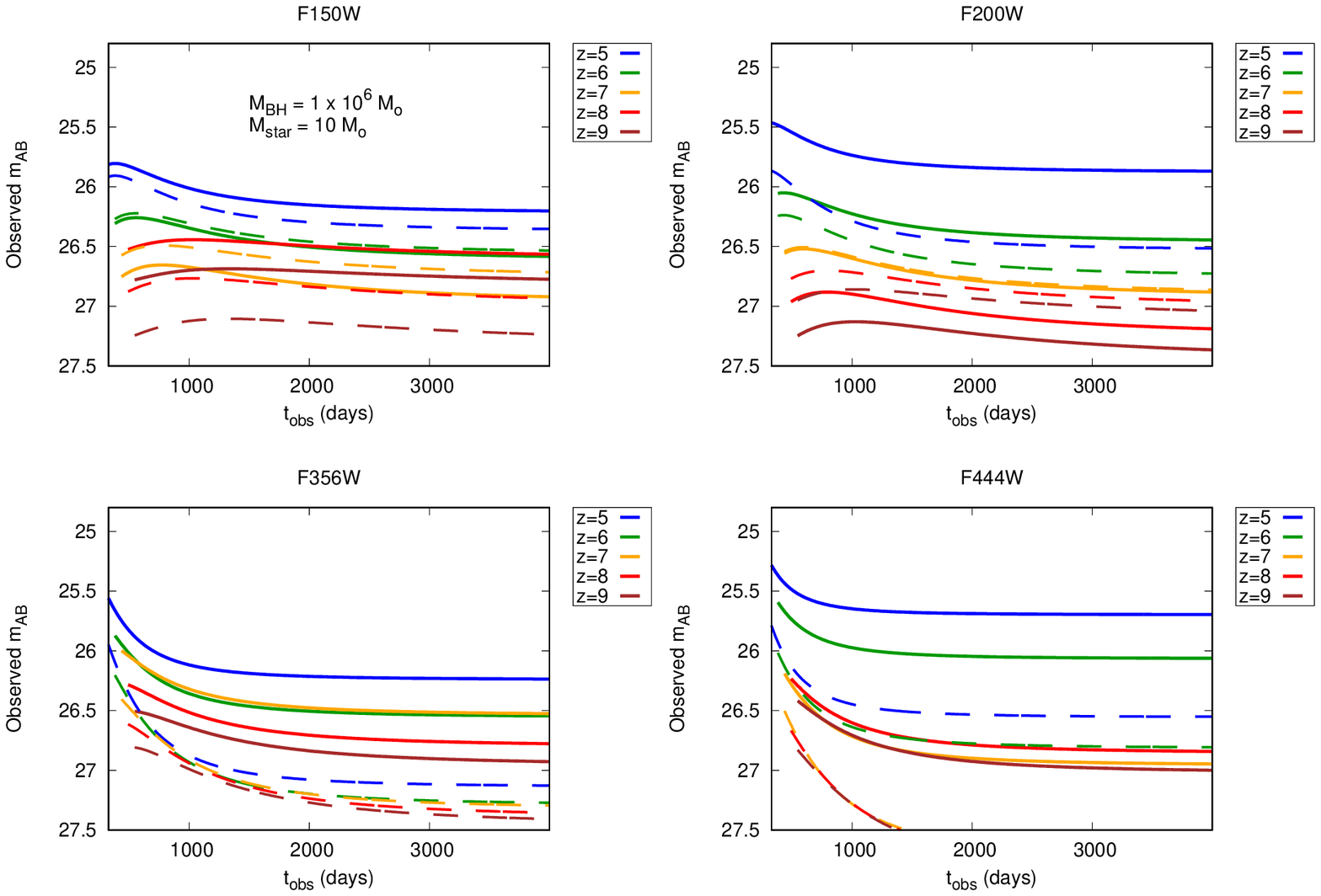}
    \caption{The same as Fig.\ref{fig:tde6200} but for $M_{star} = 10$ M$_\odot$.}
    \label{fig:tde610}
\end{figure*}

Figures~\ref{fig:tde650} and \ref{fig:tde610} are the same as Figure~\ref{fig:tde6200}, but for a 50 \Msol\ and 10 \Msol\ star, respectively.

\begin{figure*}
    \centering
    \includegraphics[width=18cm]{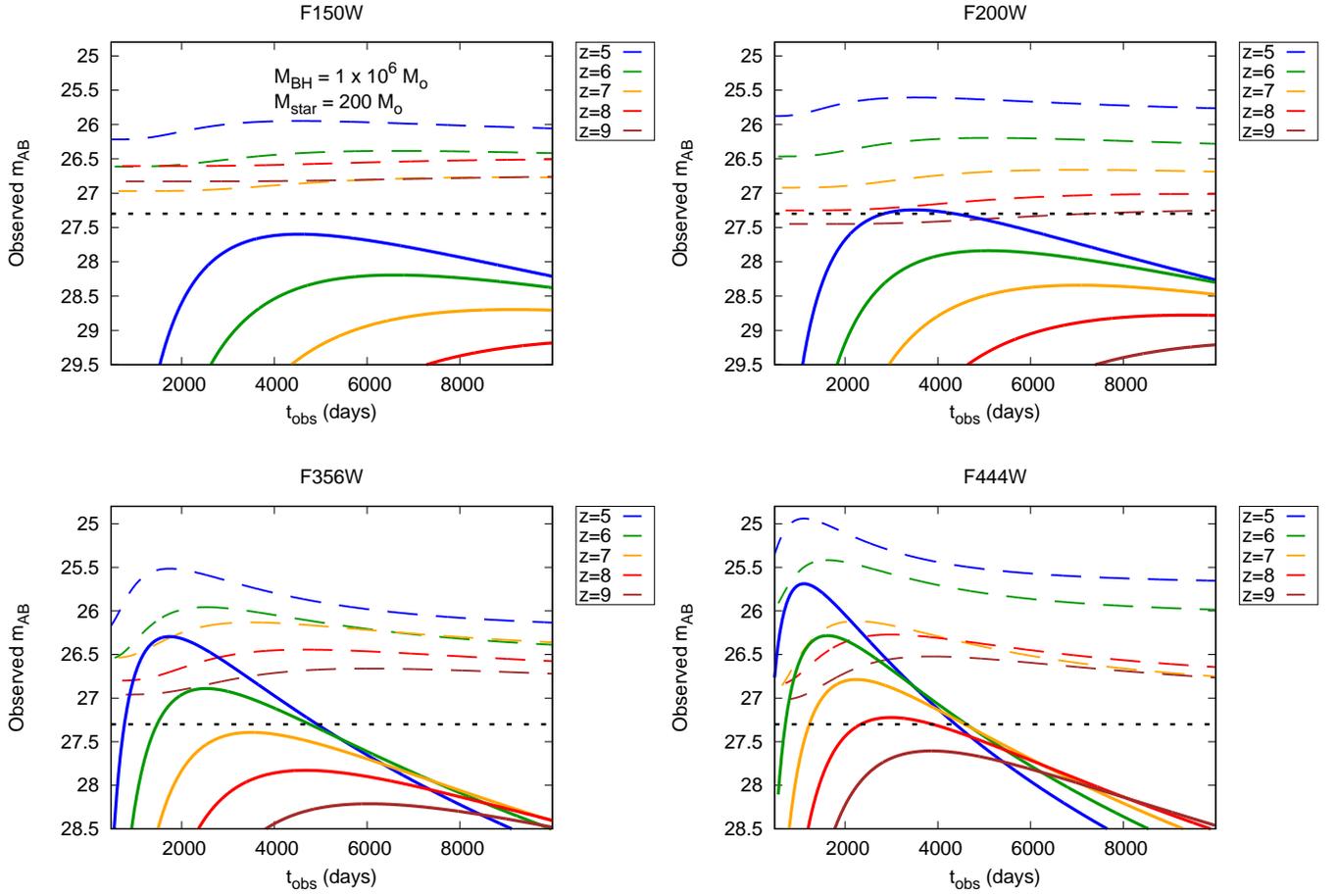}
    \caption{{ The same as Fig.\ref{fig:tde6200} but for non-accreting TDE, i.e. without the DCBH accretion SED contribution (solid lines). For comparison, the combined DCBH+TDE light curve assuming standard accretion (Fig.\ref{fig:tde6200}) is also shown with dashed lines. Dotted horizontal line denotes the proposed detection limit of the $FLARE$ survey.}}
    \label{fig:tde6200-2}
\end{figure*}

\begin{figure*}
    \centering
    \includegraphics[width=18cm]{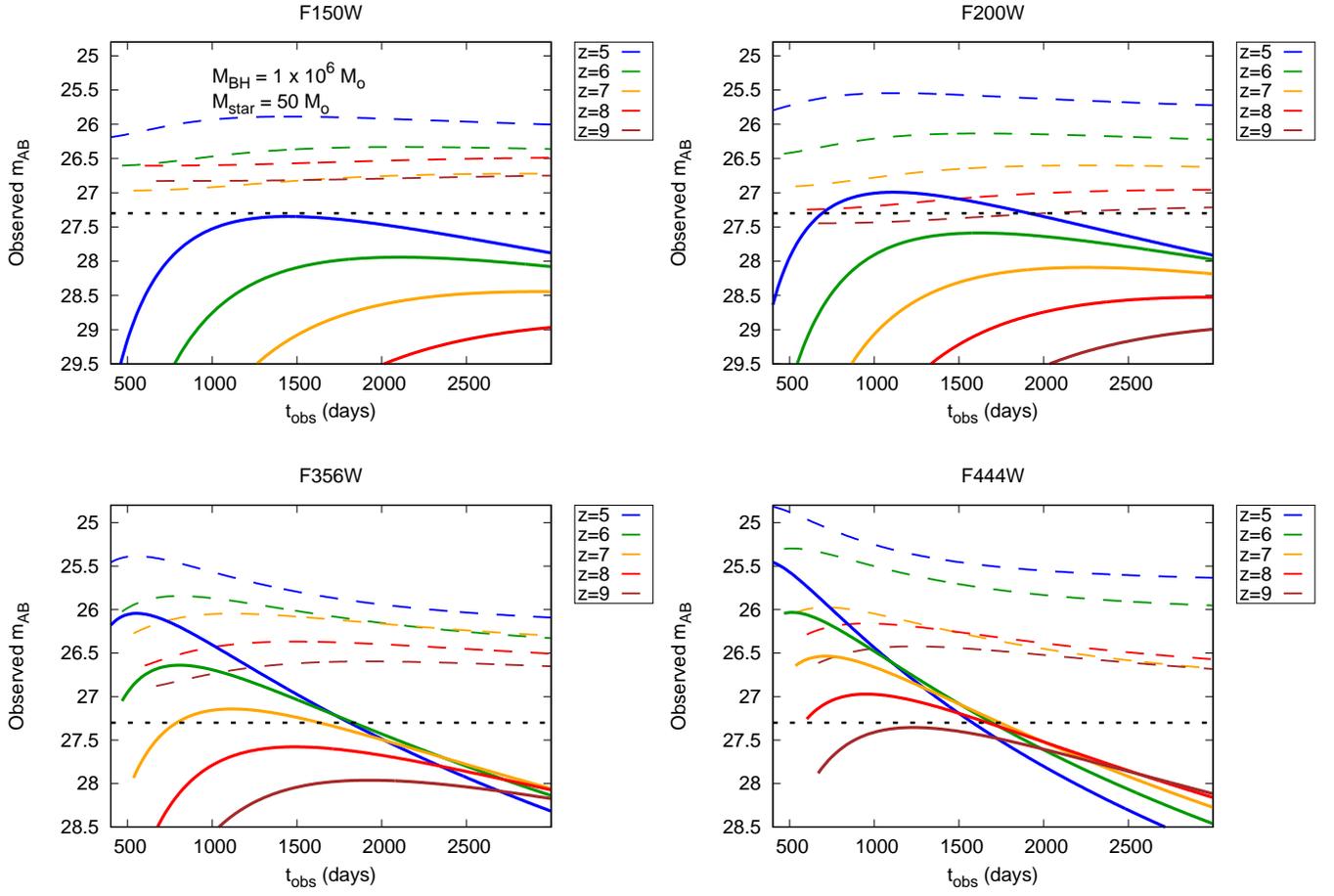}
    \caption{{ The same as Fig.\ref{fig:tde6200-2} but for $M_* = 50$ \Msol.}}
    \label{fig:tde650-2}
\end{figure*}

\begin{figure*}
    \centering
    \includegraphics[width=18cm]{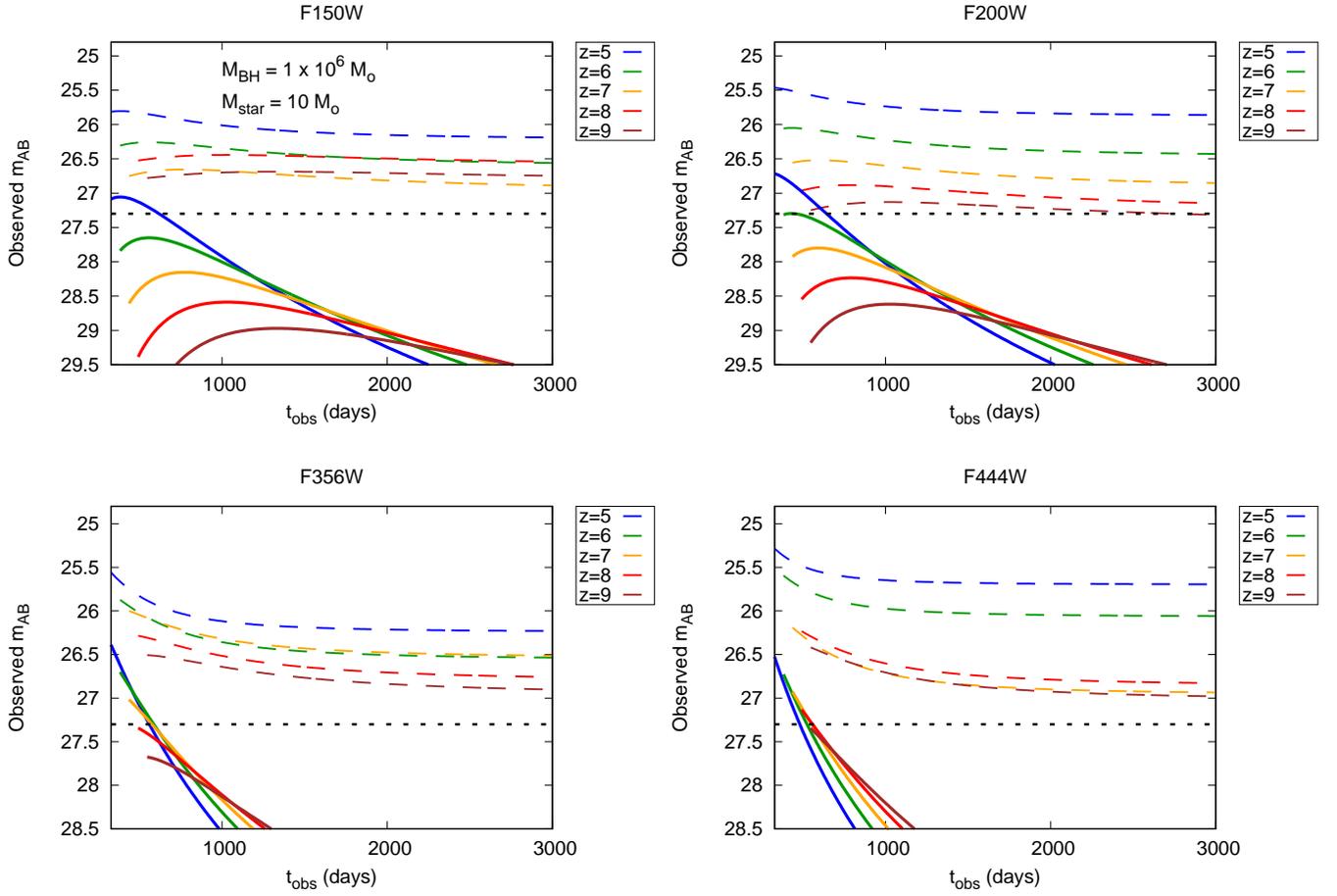}
    \caption{{ The same as Fig.\ref{fig:tde6200-2} but for $M_* = 10$ \Msol.}}
    \label{fig:tde610-2}
\end{figure*}

{ The solid lines in} Figures~\ref{fig:tde6200-2} -- \ref{fig:tde610-2} display the light curves of TDEs from 200, 50 and 10 \Msol\ stars around a $10^6$ \Msol\ black hole without the DCBH accretion luminosity contribution. { The combined DCBH + TDE signal assuming standard accretion (Figures~\ref{fig:tde6200}-\ref{fig:tde610}) is also shown with dashed lines for comparison.}



{ From Fig.~\ref{fig:tde6200}, \ref{fig:tde650} and \ref{fig:tde610}} it is seen that a TDE signal on top of the DCBH accretion luminosity { may} enhance the detectability of such events with JWST. In the $5 < z < 7$ redshift range all { the combined DCBH + TDE} curves stay above the 27 mag detection limit, regardless of the assumed accretion model, while between $7 < z < 9$ only the TDE+DCBH models with standard accretion pass the detection limit. The mass of the disrupted star affects mostly the timescale of the transient event. In the case of a (hypothetical) 200 \Msol\ star (Fig.\ref{fig:tde6200}) { the signal shows significant ($\gtrsim 0.5$ mag)} change during 1000 days (the assumed duration of the {\it FLARE} survey) in the two redder bands (in $F356W$ and $F444W$), { while in the $F150W$ and $F200W$ bands the amplitude of the variation due to the TDE is only $\lesssim 0.1$ mag.}  Such a massive star could be either a massive Population II or Population III (metal-free) star before its final fate as a PISN \citep{rvz20}.

For the disruption of the 50 \Msol\ and 10 \Msol\ star (Fig. \ref{fig:tde650} and \ref{fig:tde610}) the predicted timescales are shorter, { but the variation stays more pronounced (i.e. $\gtrsim 0.5$ mag)} in the two redder bands { during the first 1000 days.} 

On the other hand, Fig. \ref{fig:tde6200-2}, \ref{fig:tde650-2} and \ref{fig:tde610-2} indicate that detection of TDEs around non-accreting $M\sim 10^6$ \Msol\ SMBHs that no longer show the DCBH accretion luminosity is feasible only with $F356W$ at $z<6$ and with $F444W$ in the $z<7$ redshift range. The disruption of a 200 \Msol\ star produces a detectable, but slowly varying signal in these bandpasses and redshifts. The detectability of the TDE of a 50 \Msol\ star looks more promising, since the peak magnitude is similar but the timescale is shorter, which may help the identification of the varying signal during the transient survey. The TDE of a 10 \Msol\ star { around a non-accreting SMBH} (Fig. \ref{fig:tde610-2}) looks just marginally detectable for a short period of time ($t < 1$ year). 

\begin{figure*}
\centering
\includegraphics[width=8.5cm]{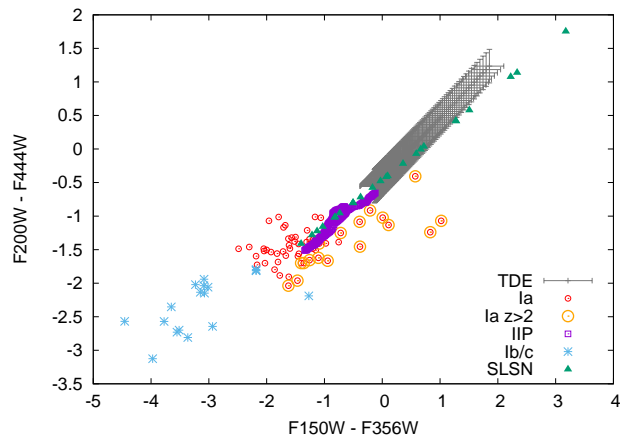}
\caption{Color-color plot for TDEs { from non-accreting SMBHs} (grey symbols) observed with NIRCam filters. The colored symbols indicate various types of supernovae as shown in the legend \citep{rv19}. { Due to their higher redshifts, TDEs appear redder than other transients during the early phases, but they get bluer as they evolve.}}
\label{fig:tdecc}
\end{figure*}

Figure~\ref{fig:tdecc} shows the location of TDEs { from non-accreting BHs} (grey band) on the NIRCam color-color plot similar to Fig. \ref{fig:dcbh_cc}. It is seen that such events occupy almost the same range as superluminous supernovae (SLSNe) \citep{rv19}. This makes the separation of these two types of transients difficult, since the timescales of SLSNe are also long, comparable to the slow evolution of massive TDEs. { Note that during their early phases TDEs appear redder than most of the other transients in this plot, because of their higher redshifts ($z > 4$), but they become bluer as they evolve, moving closer to the observable color regime occupied by SNe Ia and II-P.} 

\begin{figure*}
\centering
\includegraphics[width=8.5cm]{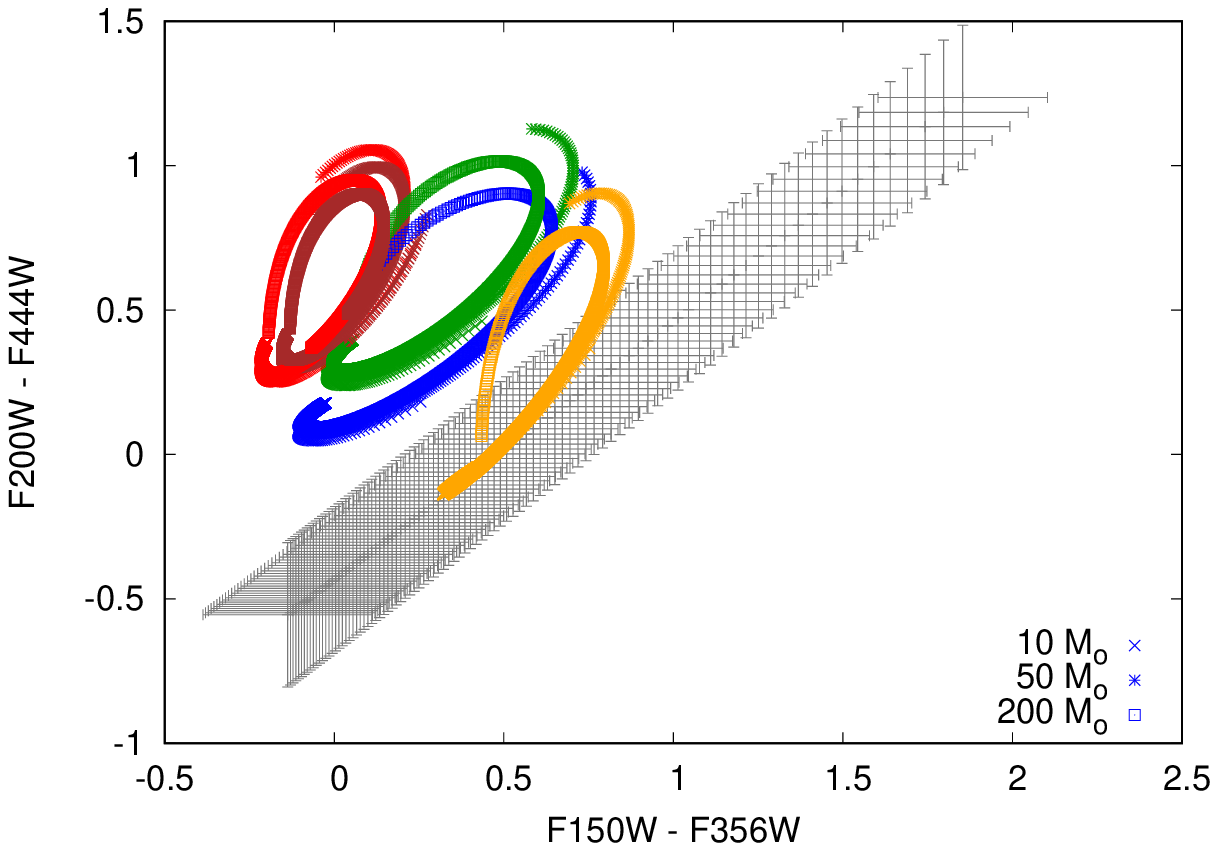}
\includegraphics[width=8.5cm]{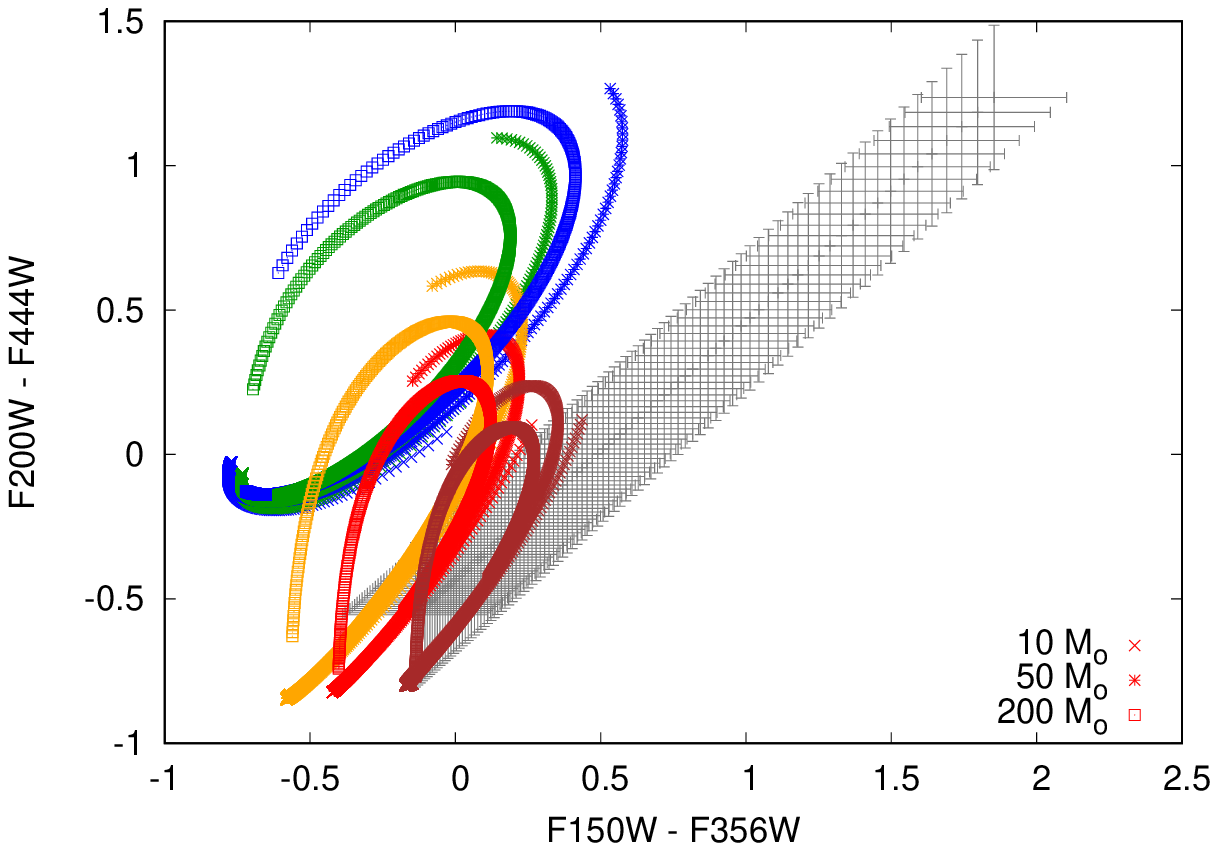}
\caption{Color-color plot of TDEs with DCBH accretion. Different colors code different redshifts similar to Fig. \ref{fig:tde6200}-\ref{fig:tde610}. Symbols correspond to stellar masses in M$_\odot$ as shown in the legend. Left panel: standard accretion model. Right panel: slim-disk accretion model. The grey band indicates the expected location of TDEs without DCBH accretion contribution (see Fig.\ref{fig:tdecc}). }
\label{fig:tdecc23}
\end{figure*}

As a comparison, Figure~\ref{fig:tdecc23} displays the location of TDE+DCBH signals on the same color-color plot (again, the grey band marks the range of the pure TDEs). Here the same color-coding scheme has been applied as above in Fig.\ref{fig:tde6200}--\ref{fig:tde610}. Like in Fig.\ref{fig:dcbh_cc}, the accretion onto the DCBH moves the measured colors of such events to the upper left part of the color-color plot, but the time-dependence of the signal make them more dispersed with respect to the pure DCBH accretion plotted in Fig.\ref{fig:dcbh_cc} (the loop-like structures could be just due to the approximate TDE light curves, though, and in the real data they might not appear the same as shown here).  

\section{Conclusions}
We studied the detectability of Tidal Disruption Events in the vicinity of either Direct Collapse Black Holes or Supermassive Black Holes with JWST NIRCam during the proposed $FLARE$ transient survey \citep{flare}. We combined models for the accretion luminosity of DCBHs with different rates from \citet{pacucci15} with analytic TDE light curves of \citet{lr11}, redshifted and convolved with the NIRCam filter bandpasses by applying the {\tt sncosmo} code \citep{sncosmo}.

Based on the results presented in the previous sections, we draw the following conclusions: 
\begin{itemize}
    \item{DCBH models with standard accretion offer the possibility of detection of such seeds with NIRCam for $t \lesssim 110$ Myr after collapse in all bands at $z < 7$ redshifts. However, DCBH models with slim-disk accretion seem to be too faint for the detection limit of the survey ($\sim 27$ AB-mag). These objects are shifted upward on the NIRCam color-color diagram, separated from the range occupied by supernovae.}
    \item{TDE around DCBH seeds may enhance the detectability of such events as a time-dependent signal. TDEs may make the DCBH seeds detectable up to $z \sim 9$ in the $F356W$ and $F444W$ bands.}
    \item{TDEs around SMBHs without the DCBH accretion luminosity are less bright than DCBHs, but still might be detected at $z<7$ in the two redder NIRCam bands. The disruption of intermediate mass $M \sim 50$ \Msol\ stars offer the best opportunity for detection, regarding both peak luminosity and variation timescale during the survey time. }
    \item{Such TDEs appear close to the range of SLSNe on the NIRCam color-color plot, which may make them difficult to distinguish from supernovae.}
\end{itemize}

\acknowledgments
We are grateful to Lifan Wang, Jeremy Mould and all other members of the FLARE team for enlightening discussions.
{ We thank Fabio Pacucci for providing DCBH spectra, and the Referee for helpful comments and insight.}
This work has been supported by the project "Transient Astrophysical Objects"  GINOP 2.3.2-15-2016-00033  of the National Research, Development and Innovation Office (NKFIH), Hungary, funded by the European Union.

\vspace{3mm}
\facilities{JWST}

\software{astropy \citep{2013A&A...558A..33A},  
          sncosmo \citep{sncosmo}}

\end{document}